\documentclass[11pt]{article}
\usepackage{fullpage}
\usepackage{amssymb}
\usepackage{amsmath,amsthm}
\usepackage{multirow}
\usepackage{color}
\usepackage{xspace}

\usepackage{ifpdf}
\ifpdf    
usepackage{hyperref}
\else    
\usepackage[hypertex]{hyperref}
\fi

\newcommand{\remove}[1]{}

\newtheorem{theorem}{Theorem}[section]

\newtheorem{cor}{Corollary}[section]
\newtheorem{lemma}[theorem]{Lemma}

\newcommand{\comment}[1]{}
\newcommand{\junk}[1]{}
\newcommand{\conf}[1]{}
\newcommand{\full}[1]{}

\begin{document}
\title{On-line Indexing for General Alphabets \\via Predecessor Queries on Subsets of an Ordered List}



\author{ Tsvi Kopelowitz\thanks{This work was supported in part by The Israel Science Foundation (grant \#452/08), by a US-Israel BSF grant \#2010418, and by the Citi Foundation.} \\ Weizmann Institute of Science \\ \texttt{tsvi.kopelowitz@weizmann.ac.il}}


\maketitle

\setcounter{page}{1}

\begin{abstract}

The problem of \textit{Text Indexing} is a fundamental algorithmic problem in which one wishes to preprocess a text in order to quickly locate pattern queries within the text. In the ever evolving world of dynamic and on-line data, there is also a need for developing solutions to index texts which arrive on-line, i.e.~a character at a time, and still be able to quickly locate said patterns. In this paper, a new solution for on-line indexing is presented by providing an on-line suffix tree construction in $O(\log \log n + \log\log |\Sigma|)$ worst-case expected time per character, where $n$ is the size of the string, and $\Sigma$ is the alphabet. This improves upon all previously known on-line suffix tree constructions for general alphabets, at the cost of having the run time in expectation.

The main idea is to reduce the problem of constructing a suffix tree on-line to an interesting variant of the order maintenance problem, which may be of independent interest. In the famous order maintenance problem, one wishes to maintain a dynamic list $L$ of size $n$ under insertions, deletions, and order queries. In an order query, one is given two nodes from $L$ and must determine which node precedes the other in $L$. In an extension to this problem, named the {\em Predecessor search on Dynamic Subsets of an Ordered Dynamic List problem} (POLP for short), it is also necessary to maintain dynamic subsets $S_1,\cdots ,S_k\subseteq L$, such that given some $u\in L$ it will be possible to quickly locate the predecessor of $u$ in $S_i$, for any integer $1\leq i \leq k$. This paper provides an efficient data structure capable of locating the predecessor of $u$ in $S_i$ in $O(\log \log n)$ worst-case time and answering order queries on $L$ in $O(1)$ worst-case time, while allowing updates to $L$ in $O(1)$ worst-case expected time and updates to the subsets in $O(\log \log n)$ worst-case expected time. This improves over a previous data structure which may be implicitly obtained from Dietz~\cite{Dietz89}, in which the updates to the sets and $L$ are done in $O(\log \log n)$ \textit{amortized} expected time. In addition, the bounds shown here match the currently best known bounds for predecessor search in the RAM model.

Furthermore, this paper improves or simplifies bounds for several additional applications, including fully-persistent arrays, the monotonic list labeling problem, and the Order-Maintenance Problem.

\end{abstract}

\section{Introduction}\label{sec:intro}

{\em Text Indexing} is one of the most important paradigms in text
searching. The idea is to preprocess a text $T$ of size $n$ and construct a
mechanism that will later provide answers to queries of the
form "does a pattern $P$ of size $m$ occur in the text?" in time proportional
to $m$ rather than $n$. This paradigm can be seen in many applications of the computing world, including web searching, computational biology applications, stock market predictions, and indexing astronomical data. In the stock market scenario, for example, one also needs to consider the very large size of the alphabet $|\Sigma|$ from which the text is drawn. Such a consideration also needs to be taken when dealing with astronomical data, such as indexing the sky~\cite{SGKTSB00}. The suffix tree and suffix array have proven to be invaluable data structures
for indexing.

One of the problems that has occupied the algorithmic
community is that of constructing an on-line or real-time indexing algorithm. An
algorithm is \emph{on-line} if it accomplishes its task for the $i$th
input without needing the $i + 1$st input. It is \emph{real-time} if,
in addition, the time it operates between inputs is a constant.

Some of the static suffix tree constructions work in the {\em on-line} model~\cite{Ukkonen95,Weiner73}, in which one maintains a suffix tree for a text that arrives character by character, and at any given time one might receive a pattern query. For simplicity sake, assume that the text arrives from the end towards the beginning. This is because it can be shown that a single character added at the end of the text can impose a linear number of changes to the suffix tree. Of course, if the text arrives from beginning to end, one can view the text in reversed form, and then a queried pattern is reversed as well in order to obtain the correct results. The currently best known results for the on-line suffix tree construction for general alphabets are an $O(\log |\Sigma|)$ amortized time per character by Weiner in~\cite{Weiner73}, and $O(\log n)$ worst-case per character by Amir, Kopelowitz, Lewenstein, and Lewenstein in~\cite{AKLL05}, where $n$ is the size of the text and $\Sigma$ is the alphabet. As far as constant sized alphabets are concerned, Breslauer and Italiano in~\cite{BI11} recently obtained an algorithm which costs $O(\log \log n + |\Sigma|)=O(\log \log n)$ worst-case time per character. For constant-size alphabets there is another indexing structure by Amir and Nor~\cite{AN08} which does not enjoy the various advantages of the suffix tree.

The results here beat the cost per character of all of the above algorithms for general alphabets. However, the time bounds are in expectation (but not amortized). It is proven here that a suffix tree can be updated on-line in $O(\log \log n + \log \log |\Sigma|)$ worst-case expected time per character. This is under the natural assumption that each character in $\Sigma$ fits into a constant number of words in memory. The idea behind the approach used here, as shown in section~\ref{sec:ST}, is to reduce the on-line suffix tree construction to a new data structure problem called the {\em Predecessor Search on Dynamic Subsets of an Ordered Dynamic List Problem} (POLP). This problem is defined and discussed in detail next.

\subsection{Predecessor Search on Dynamic Subsets of an Ordered Dynamic List}

The POLP, in a simplistic view, is a combination of two well studied problems: the {\em Order-Maintenance Problem} (OMP) and the Predecessor problem.

\paragraph{Order-Maintenance:} In the OMP the goal is to maintain an ordered list $L$ of size $n$ under the following operations: (1) $Insert(u,v)$ - Insert node $u$ after node $v$ in $L$. (2) $Delete(u)$ - Delete node $u$ from $L$. (3) $Order(u,v)$ - Determine whether $u$ preceeds $v$ in $L$.

The first data structure to solve the OMP was introduced by
Dietz in~\cite{Dietz82}, which was later improved together with Sleator in~\cite{DS87}, where in the first part of their paper they introduced a solution to a different problem known as the Monotonic
List Labeling Problem (MLLP). In the MLLP the goal is the same as in the OMP with the additional constraint that each node in $L$ is given a unique tag
from a range bounded polynomially in $n$. Then, ordered queries are performed by comparing tags in constant time. Dietz and Sleator in~\cite{DS87}
showed how the MLLP can be solved such that each insertion costs $O(\log n)$ amortized time. They then used this solution to provide a solution for the OMP where each insertion costs $O(1)$ amortized time. About 15 years later, Bender, Cole, Demaine, Farach-Colton, and Zito in~\cite{BCDFZ02} simplified both of these solutions, obtaining the same time bounds. They further claimed that their amortized solution can be made worst-case, and deferred the description to the full paper, which unfortunately has yet to be published. As mentioned by Bender, Cole, Demaine, Farach-Colton, and Zito in~\cite{BCDFZ02}, the results of Dietz and Sleator in~\cite{DS87} rely on a complicated counterintuitive potential-function in order to achieve their amortized cost solution (This is cited by Dietz and Sleator~\cite{DS87} as a disadvantage of their analysis). Dietz and Raman in~\cite{DR93} obtained another solution for the MLLP, where each insertion relabels $O(\log n)$ tags in the worst-case, in $O(\log^2 n)$ worst-case time. They further claim that it is possible to reduce the time to $O(\log n)$ worst-case, but many non-trivial details are missing. A lower bound showing that at least $\Omega(\log n)$ relabelings need to take place per insertion was shown by Dietz, Seiferas, and Zhang in~\cite{DSZ94}.

A third closely related problem, called the File-Maintenance Problem (FMP), is the same as the MLLP, except now the range of the tags is bounded by $O(n)$. Willard in~\cite{Willard86} was the first to provide a very complicated solution for the FMP, where each insertion costs $O(\log^2 n)$ worst-case time. Bender, Cole, Demaine, Farach-Colton, and Zito in~\cite{BCDFZ02} gave a simpler solution for the FMP, but some proofs are missing. Recently, Bul{\'a}nek, Kouck{\'y}, and Saks in~\cite{BkKS12} showed a matching lower bound. In the second part of~\cite{DS87}, Dietz and Sleator showed how Willard's complicated solution for the FMP can be used to provide a solution for the OMP, where each insertion costs $O(1)$ worst-case time.

Alas, to date there is no published version of a worst-case solution for the OMP which is not considered highly complex. This is especially surprising considering that the OMP is a common building block for many data structures in dynamic settings. Furthermore, for the purposes of the results presented here, the solution of Dietz and Sleator~\cite{DS87} does not suffice, as is explained later.

\paragraph{Predecessor Queries:} Predecessor data structures are ubiquitous in the computer science
literature. The {\em static predecessor} problem is to store a set
of integers and perform predecessor queries on the set. The {\em
dynamic predecessor} problem also allows insertions and deletions on
the set. In the comparison based model an $\Omega(\log n)$ lower bound for the predecessor search within a set of size $n$ is easy to obtain. However, improved bounds are possible in the RAM model\footnote{All of the results in this paper are in the RAM model.}.

Several data structures have been proposed for the predecessor
problem. For example, the van Emde Boas data structure~\cite{vEB77} was presented as an efficient data structure
for small universes. Namely, operations are performed in $O(\log\log
u)$ time for universe of size $u$. The space for the van Emde Boas
data structure is $O(u)$, and can be reduced to $O(n)$ using randomization. Other solutions, such as
x-fast tries and y-fast-tries~\cite{Willard83}, have been suggested as well.
\conf{Fredman and Willard broke the $\Omega(\log n)$ lower bound for the
predecessor queries with the fusion tree~\cite{FW93}.
}
There are
many other results and the interested reader is directed
to~\cite{BF02,PT06} for other upper and lower bounds on the
problem.

\paragraph{Combining the Two:} In the POLP, the goal is to maintain a dynamic ordered list $L$, but, in addition to order queries, there are dynamic subsets $S_1,\cdots,S_k\subseteq L$ which need to be supported to answer predecessor queries. In a predecessor query on a set $S_i$, the input is a node $u\in L$ and the output is the largest element $v\in S_i$ which is smaller than $u$, where the order is defined by $L$. For simplicity sake, assume that the subsets are disjoint. An exposition of the case of non-disjoint sets is left for the full paper.

The goal is to support the operations on the subsets in time which is proportional to the size of $L$, and not dependent on the size of the universe from which the elements in $L$ are drawn from. In this sense, the POLP setting can be viewed as some sort of embedding. However, this will only work efficiently if it is possible to quickly locate the position in $L$ of every new element, which is not part of the scope of this paper. Given how useful the order-maintenance data structure has been in the data structure world, it is entirely conceivable that other than the on-line suffix tree construction many more applications exist for the POLP.

The POLP was first implicitly solved by Dietz in~\cite{Dietz89} where a solution for fully persistent arrays was introduced. The structure there answers order queries in $O(1)$ worst-case time, performs insertions into $L$ and the subsets in $O(\log \log n)$ expected amortized time, and answers predecessor queries in $O(\log \log n)$ time. However, the solution there does not deal with deletions.

In this paper, the focus is on developing a worst-case solution. Thus, the data structure presented here performs insertions and deletions to the list in $O(1)$ worst-case expected time and to the subsets in $O(\log \log n)$ worst-case expected time. Answering order queries is done in $O(1)$ worst-case time and answering predecessor queries is done in $O(\log \log n)$ worst-case time.

Due to space limitations, the discussion of deletions is deferred to the full version\footnote{Notice that for the OMP deletions can be easily dealt with by marking nodes as deleted, and using a standard rebuilding technique once the number of deleted nodes in $L$ becomes large enough. Also notice that one cannot delete a node $u\in L$ if it is in some set.}. Nevertheless, for the purposes of the applications mentioned here, deletions are not needed. Notice that these time bounds match the currently best known bounds for the dynamic predecessor problem. It is currently a very interesting open problem whether or not the expectation can be removed from the update time in the dynamic predecessor problem, without increasing the query time (\cite{Patrascuprivate}).

\subsection{The Difficulties}
Dietz and Sleator in~\cite{DS87} in their complicated solution for the OMP provide each element in $L$ with at most two tags, each with a timestamp, such that given two nodes in $L$, their order can be determined from the tags alone. The tags are integers from a range polynomial in the size of $L$. This gives some intuition as to why one might expect that solving POLP can be done within the claimed time bounds. However, there are two main difficulties that need to be dealt with in order to solve the POLP efficiently.

The first is that given how the predecessor data structures use the bit presentation of integers, it is not clear how double tags with timestamps could be made to work. The second difficulty is that an insertion of a new node into $L$ can cause $\Omega(polylog(n))$ tags of elements in $L$ to change, which could be costly for a predecessor data structure used directly on the tags.

The first problem is solved by presenting a new solution for the MLLP problem (see Section~\ref{sec:OLL}) where each insertion costs $O(\log n)$ worst case time\footnote{It is highly conceivable that Bender, Cole, Demaine, Farach-Colton, and Zito in~\cite{BCDFZ02} had a similar solution in mind when they claimed, without proof, that their amortized solution can be deamortized.}. This solution for the MLLP is then used to provide a new data structure for the OMP with worst-case bounds, where each element has only one integer tag at a given time. This is explained in more detail in Section~\ref{sec:OMP}. The second problem is solved by tying together the indirections used for the order-maintenance data structure and the predecessor data structures, together with careful scheduling of processes. This is explained in more detail in Section~\ref{sec:POLP}.

\subsection{Fully Persistent Arrays}
Due to space limitations, some more applications of the POLP are deferred to the full paper. Nevertheless, it is briefly pointed out here that by replacing Dietz's solution in~\cite{Dietz89} with the solution presented here as a black-box, the amortized time bounds of fully-persistent arrays become worst-case (though insertions are still in expectation), which immediately implies improved bounds for the general method of making any data structure fully persistent in the RAM model, as described by Dietz in~\cite{Dietz89}.

\section{On-line Suffix Tree Construction}\label{sec:ST}

In the on-line suffix tree construction, the goal is to support extensions of the text $T$ in which new characters are added to its beginning, i.e., constructing the suffix tree of $\sigma T$ from the suffix tree of $T$, where $\sigma \in \Sigma$. When referring to the suffix tree the intention is the suffix tree of $T$ before the additional character is added, unless mentioned otherwise.

The discussion here assumes the reader is familiar with the basics of the suffix tree data structure. Recall that each node in the
suffix tree has a maximum out-degree of $|\Sigma|$ (every outgoing edge represents a character from $\Sigma$, and any two outgoing
edges represent different characters). For the purposes here, a hash function is used to map each character to its appropriate edge. In addition, for any node $u$ in the suffix tree, the length of the string corresponding to the path from the root to $u$ is denoted by $\textit{length}(u)$, and is maintained within $u$. In addition, all of the suffixes are maintained within a lexicographically ordered list of suffixes.

The process of inserting the new suffix $\sigma T$ into the suffix tree is broken into three phases. The first phase locates the position of the new suffix in the list of sorted suffixes. The second phase locates the place in the suffix tree to which the new suffix needs to be added. Finally, in the third phase, insertion of the new suffix is implemented by either adding a new leaf as a child of a node already in the suffix tree, or by splitting an edge in the suffix tree into two by adding a new node $u$ into the edge, and then the new leaf is a child of $u$. In either case the machinery used needs to be updated as well.

\subsection{Phase 1: Searching in the List}

At first glance, maintaining the list of ordered suffixes in a POLP data structure seems to suffice, as all that needs to be done is perform a predecessor query on the list of ordered suffixes with the new suffix as the key. However, this will not work as the new suffix is not yet in the suffix list, while the POLP assumes that the key is already part of the list.

To solve this, notice that when comparing two different suffixes (or strings for that matter), it is possible to break down the comparison process into two. The first comparison is done by comparing the first character. If the first two characters are different, then the order of the two suffixes is determined by just those characters. Otherwise, the rest of those two suffixes will set the order. So when attempting to locate the predecessor of $\sigma T$: (1) locate the consecutive list in the ordered list of suffixes of $T$ which all correspond to suffixes starting with $\sigma$, and (2) within this sublist, find the predecessor of $\sigma T$.

To solve (1) efficiently one can use any predecessor data structure on the different characters of $\Sigma$ which appear in $T$. Using a y-fast-trie structure~\cite{Willard83}, for example, allows to locate the sublist in $O(\log \log |\Sigma|)$ time. Notice that a hash function will not suffice here as there is a need to know the order between the different characters present in $T$, and it is possible that this is the first time $\sigma$ appears.

To solve (2), notice that the order of the suffixes in the sublist corresponding to suffixes of $T$ beginning with $\sigma$ is determined by truncating $\sigma$ from each of those suffixes, and determining the order of the remaining substrings. Luckily, each of those substrings is also a suffix of $T$. So for each $\sigma\in \Sigma$, a predecessor structure $P_{\sigma}$ is maintained over the nodes from the suffix list which correspond to suffixes that begin with $\sigma$, where the keys are the truncated suffixes. In other words, the key for each suffix $\sigma T'$ in this set is the node of suffix $T'$ in the order-maintenance structure. Notice that when truncating $\sigma T$, the remaining suffix $T$ is also in the ordered suffix list, and so performing a predecessor query on $P_{\sigma}$ where the key being searched is the node in the suffix list corresponding to $T$ will find the location of the predecessor of the suffix $\sigma T$ in the ordered list. It will be shown in Section~\ref{sec:POLP} that such a query will cost $O(\log \log n)$ worst-case time.

\subsection{Phase 2: Searching in the Tree}
Being that the techniques used in this phase are either standard or use other data structures as a black box, only a sketch of the process is presented.

Once the location in the list of ordered suffixes is found, it is time to locate the place in the suffix tree into which the leaf of the new suffix needs to be added. The insertion of the new suffix is implemented by either adding a new leaf as a child of a node already in the suffix tree, or by splitting an edge in the suffix tree into two by adding a new node $u$ into the edge, and then the new leaf is a child of $u$. In either case, notice that this entry point is on the path from the root of the suffix tree to one of the neighbors of $\sigma T$ in the list of ordered suffixes. To determine which of the neighbors is the one of interest one can perform a \textit{Longest Common Prefix} (LCP) query in constant time, using the data structure of Franceschini and Grossi in~\cite{FG04}
\footnote{Similar to phase 1, the LCP of two strings can be determined by either the first character, or the LCPs of the suffixes without the first character.}
. Then, one can locate the entry point in $O(\log \log n)$ time using weighted level ancestor queries on dynamic trees~\cite{KL07}, where the weight of each node is its length, and the query is the LCP of $\sigma T$ and its appropriate neighbor. The entire process takes $O(\log \log n)$ time.

\subsection{Phase 3: Updating the Suffix Tree and Machinery}

If the new suffix is inserted as a child of a pre-existing node, then the new edge leading to the new leaf is inserted into the appropriate hash function used for navigation down the suffix tree. The new leaf needs to be inserted into the machinery used (i.e. weighted level ancestor queries). This can be done in $O(\log \log n)$ worst-case expected time~\cite{KL07}. Also, the new suffix needs to be inserted into the LCP data structure, which takes constant worst-case time~\cite{FG04}, and if $\sigma$ is a new character in the text it needs to be inserted into the vEB structure for the alphabet in $O(\log \log |\Sigma|)$ worst-case expected time. Finally, the new suffix is added into the suffix list and the node in the suffix list corresponding to $T$ is added to $P_{\sigma}$, both of which are done by updating the POLP data structure. It will be shown in Section~\ref{sec:POLP} that such an update will cost $O(\log \log n)$ worst-case expected time.

If the new suffix is inserted together with a new inner node then the inner node needs to update the hash of its parent, create a hash function for itself containing the end of the edge it broke (i.e. the previous child of the new inner node's parent), and be inserted into the machinery used on the trees. This can be done in $O(\log \log n)$ worst-case expected time~\cite{KL07}. The insertion of the new leaf is performed as before.

Notice that many of the operations on suffix trees (assuming linear space is desired) use various pointers to the text in order to save
space for labeling the edges. It is shown by Amir, Kopelowitz, Lewenstein and Lewenstiein in~\cite{AKLL05} how to maintain such
pointers, called \emph{text links}, within the time and space
constraints. Also notice that a copy of the text saved in
array format may be necessary for various operations, requiring direct
addressing. This can be done with constant time
update by standard de-amortization techniques. Thus, the following is obtained.

\begin{theorem}
There exists an on-line suffix tree construction where the cost for each addition of a character is $O(\log \log n + \log \log |\Sigma|)$ worst-case expected time.
\end{theorem}

\section{Monotonic List Labeling}\label{sec:OLL}
Following the methods of both Dietz and Sleator in~\cite{DS87} and Bender, Cole, Demaine, Farach-Colton, and Zito in~\cite{BCDFZ02}, each element in $L$ is provided with a tag, such that given two nodes in $L$, their order can be determined from their tags alone. For the purpose of this paper, a worst-case implementation is needed that can support the needs of the predecessor data structures which will be used in Section~\ref{sec:POLP}.

\junk{In addition, as mentioned above, the claim of the existence of a simple Order-Maintenance data structure made by Bender, Cole, Demaine, Farach-Colton, and Zito in~\cite{BCDFZ02} is still awaiting a published version with a proof, while the solution by Dietz and Sleator~\cite{DS87} is considered complex and in any case does not fit the needs here as discussed. Therefore, there is need for a new and clear solution for the OMP in which at any given time every element in $L$ has only one integer tag. Such a solution is developed here.}

\subsection{The Averaging Method.}
One possible tag-scheme would be to assign a number to a newly inserted node which is the average of the tags of its two neighboring nodes. The problem with this solution is that each tag would require $n$ bits, and so determining the order of two nodes would take $O(\frac{n}{\log n})$ time, and not $O(1)$ time which is the goal. Thus, a different solution is needed. Nevertheless, if the number of nodes in the list is $O(\log n)$, then this solution can indeed be used. This solution is named the \textit{averaging method}, and will be used on some small lists in the solution presented for larger lists.

\subsection{The Weight Balanced B-Tree}

Arge and Vitter in~\cite{AV03} introduced the Weight Balanced B-Tree (WBBT). In the WBBT, data is maintained in the leaves. The weight of each leaf is defined as the number of elements in that leaf, and the weight of an internal node is the sum of the weights of its children, i.e., the sum of the weights of the leaves in its subtree. The WBBT is defined as follows, for branching parameter $a>4$ and leaf parameter $k>0$:
\begin{itemize}
\item {} All of the leaves are at the same depth, and have weight between $k$ and $2k-1$.
\item {} An internal node of height $h$ has weight at most $2a ^ h k$, and every internal node, except for the root, has weight at least $\frac{1}{2}a ^ h k$.
\end{itemize}

Arge and Vitter proved the following (proof is omitted here):

\begin{lemma}[from~\cite{AV03}] Every internal node in the WBBT has between $a/4$ and $4a$ children, except for the root which has between $2$ and $4a$ children.
\end{lemma}

\begin{cor} [from~\cite{AV03}] The height of the WBBT with $n$ elements is $O(\log_a \frac{n}{k})$.
\end{cor}

For the purpose of the application here, $k$ and $a$ are some constants, and thus the height of the WBBT is $O(\log n)$. The elements in $L$ are maintained in the leaves of the WBBT. When an insertion is made, the $O(\log n)$ ancestors of the appropriate leaf are informed that their weight has increased. This might cause the size of some nodes to become too large, as their weight is above the allowed bounds by the definition of the WBBT. Such nodes are called overflowed nodes. Every overflowed node $u$ is split into two new nodes $u_L$ and $u_R$, and $u$'s children are divided as evenly as possible between the two new nodes. Each split requires constant time, for a total of $O(\log n)$ time to insert a new node. Arge and Vitter proved the following.

\begin{lemma}[taken from~\cite{AV03}]\label{lemma:separator_size} If the weight of $u$ prior to it splitting is denoted by $W$, then after the split, the weights of $u_L$ and $u_R$ are both $\Omega(W)$. Thus, at least $\Omega(W)$ insertions need to be made into the subtree of $u_L$ ($u_R$) before it must split again.
\end{lemma}

Lemma~\ref{lemma:separator_size} is a crucial and useful property of the WBBT, as it provides a method of informing $u$'s subtree that $u$ has split before another split happens to either $u_L$ or $u_R$. This is done by performing a scan of the subtrees of $u_L$ and $u_R$, which is spread over the insertions into any of those subtrees.

It is also important to notice that if the root ever has to split (due to its weight becoming too large), then a new root is created as the parent of the old root. For simplicity sake, assume without loss of generality that the root is never split. This assumption can be made because a rebuilding scheme can be used in case the weight of the root ever reaches its upper bound. Being that this is fairly standard, details are omitted.

\subsection{An $O(\log n)$ Implementation}\label{subsec:tag_process}

\paragraph{The Tag Scheme:} Denote by $H$ the height of the WBBT. Each element in $L$ is assigned a bit string of length $(4a+1)H+2k = O(\log n)$ as its tag, which is treated as an integer. Each level in the WBBT is responsible for $4a+1$ bits, except for the leaves which are responsible for $2k$ bits. The way the responsibility works is that each child $v$ of a node $u$ at level $i$ has a different bit string of length $4a+1$, denoted by $label(v)$, which is called the label of $v$. In $v$'s subtree, all of the elements (in the leaves) have $label(v)$ as a substring of their tag, at locations $[(4a+1)i+1,(4a+1)(i+1)]$. If the labels of the children of $u$ are assigned in such a way that the labels maintain the order of the children of $u$, then by comparing the tags of two elements, each in a subtree of a different child of $u$, the order of these elements will be determined by the order of the labels of those children of $u$. The reason the scheme works is because the path from the root to $u$ is the same for both elements, and so the most significant bit which differs must be related to the labels assigned to the children of $u$. The last $2k$ bits are assigned by the leaf to the elements within it, using the averaging method.

It must be guaranteed that the order of the children of $u$ is correctly represented within the label of each of the children. This is done by using the averaging method on the $4a$ least significant bits of each label. The use of the extra most significant bit is revealed later. Being that each node does not have more than $4a$ children, $4a$ bits suffice. It is important to notice that the splitting of $u$ into $u_L$ and $u_R$ is done by setting $u_L=u$, and inserting $u_R$ after $u_L$ in the list of children of the parent of $u$. In addition, the labels of the children of $u$ all need to be reassigned in order to spread them out within the range defined by $4a$ bits. This reassignment can be easily afforded as the number of children is bounded by a constant. Thus, splitting a node and updating the labels of its children takes constant time.

\subsubsection{Updating Tags from New Labels}
Once a split occurs, there is still a need to update all of the elements in the subtrees of $u_L$ and $u_R$ with the new labels. The process of this update is called a \textit{tag-process} and is denoted by $P_u$ for a process initiated by $u$ splitting. When an insertion is made into a leaf of the WBBT, the $O(\log n)$ ancestors of the appropriate leaf are informed that their weight has increased. Each time a node has its weight increased it is given 1 unit of time resource which needs to be spent immediately (so the time resources do not accumulate). The time resource given to either $u_L$ or $u_R$ is then given to $P_u$ and is used to pay for $O(1)$ operations performed by $P_u$. Of course, if $P_u$ has completed then the time resource is discarded. Notice that there will be situations in which $P_u$ will give its time resource to a different tag-process to use, as will be explained later. In any case, this time resource scheme will guarantee that the total amount of work performed by tag-processes due to an insertion into $L$ is bounded by $O(\log n)$.

Denote by $W$ the weight of $u$ prior to the split. Due to Lemma~\ref{lemma:separator_size}, at least $\Omega(W)$ insertions of elements must be made into either $u_L$'s subtree or $u_R$'s subtree before they split again. Thus, if a large enough constant number of leaves in those subtrees is updated whenever $P_u$ receives a time resource, the tags of the appropriate elements will all be updated with the new labels before the next splitting of either $u_L$ or $u_R$ occurs\footnote{There is also the issue of scanning the subtree, which is fairly standard and is done within the overall $O(W)$ work.}. It is important to notice that due to the method used in which each level in the tree is responsible for a different part of a leaf's tag, concurrent tag-processes updating labels from different levels in the WBBT do not interfere with each other.

The tag-process $P_u$ has three sequential phases. During the first phase, the subtree of $u_R$ is updated with the new label of $u_R$ replacing the previous label of $u$. This is done by first updating the rightmost leaf in the subtree of $u_R$ and ending with the leftmost leaf. The order of updates is important as to guarantee that order queries asked during the update process are answered correctly, even if the order query is performed on nodes of $L$ which are in the subtree of $u_R$. The second and third phase, which are interchangeable, are responsible for updating the subtrees of $u_L$ and $u_R$ with the new labels of the children of $u_L$ and $u_R$. For consistence sake, the second phase is assigned to $u_L$ and the third is assigned to $u_R$. The process is shown for $u_L$ as the process for $u_R$ is exactly the same.

\paragraph{Updating Labels for $u_L$'s Children:} There are several issues that need to be dealt with while updating the elements in $u_L$'s subtree, as order queries could be made during the process of updating the tags. There must be some guarantee that the tags are consistent with the true order, even if an order query is made while a tag-process isn't complete. This is the reason for the extra most significant bit within the labels. Before $P_u$ is initiated, there is a guarantee that this bit is the same for all of the labels of the children of $u$. Without loss of generality assume this bit is set to be $0$. When $u$ is split, $P_u$ reassigns labels to $u_L$'s children, but now their most significant bit is changed to 1. When the subtree of $u_L$ is updated with the new labels, it begins by updating the rightmost leaf in the subtree towards the leftmost leaf. This guarantees that if an order query is made between two elements in the subtree of $u_L$ then:
\begin{itemize}
\item{} If both elements have already been updated with the new label of $u_L$'s child, then the $4a+1$ bits for which the children of $u_L$ are responsible will correctly determine the order.
\item{} If both elements have not been updated with the new label of $u_L$'s child, then the $4a+1$ bits for which the children of $u_L$ are responsible are the same as prior to $u$ splitting, and so they correctly determine the order.
\item{} If one element, $\alpha$ has been updated, while the other element $\beta$ has not, then it must be that $\alpha$ is larger than $\beta$ (due to the order in which the leaves are updated), and so the most significant bit in the $4a+1$ bits for which the children of $u_L$ are responsible is $1$ for $\alpha$ and $0$ for $\beta$. Thus the tags correctly maintain the order.

\end{itemize}

\subsubsection{Collisions of Splitting Processes.} Let $w$ be the parent of $u$ prior to $u$ splitting, and let $v$ be a child of $u$ prior to $u$ splitting. A difficulty arises when either $w$ or $v$ split during the execution of $P_u$. If $w$ splits then care needs to be taken with regard to updating the tags in the subtree of $u_R$, as the first phase of $P_u$ will be using the label assigned to $u_R$ by $P_u$, while the third phase of $P_w$ will be using the label assigned to $u_R$ by $P_w$. If $v$ splits into $v_L$ and $v_R$ then care needs to be taken with regard to updating the tags in the subtree of $v_R$, as the second or third phase of $P_u$ will be using the label assigned to $v$ by $P_u$, while the first phase of $P_v$ will be using the label assigned to $v_R$ by $P_v$. However, it is important to notice that when $P_u$ is in a collision of this sort with $P_w$, it cannot be in a collision with $P_v$, as collisions with $P_w$ can only happen during the first phase of $P_u$, while collisions with $P_v$ only happen during the second or third phase of $P_u$. Also notice that $w$ ($v$) splitting before $u$, is analogous to $u$ splitting before $v$ ($w$).

To solve these collisions, a careful scheduling of process is needed, as is described next.

\paragraph{When $v$ splits during $P_u$:} In this case, $P_u$ assigns a new label to $v$. Then $v$ splits, and $P_v$ assigns a new label for $v_R$. At the end of the execution of both $P_u$ and $P_v$, the label assigned to $v_R$ by $P_v$ must be the label which is assigned in the appropriate locations in all of the tags of leaves in $v_R$'s subtree. This is guaranteed as follows. If $P_u$ has already updated the subtree of $v$ before $P_v$ begins updating $v_R$'s subtree, then no special modifications need to be made. If $P_v$ has already begun updating $v_R$'s subtree when $P_u$ reaches $v$'s subtree (and in particular $v_R$'s subtree), then $P_u$ uses its time resources to help $P_v$ finish updating $v_R$'s subtree. In such a case, the label assigned to $v$ by $P_u$ is never used to update any tag in $v_R$'s subtree. It is important to notice the significance of the MSB in order to understand the correctness of this process scheduling. Finally, if $P_u$ is in the process of updating $v$'s subtree when $P_v$ begins, then first $P_v$ uses its time resources to help $P_u$ finish updating $v$'s subtree with the label assigned to $v$ by $P_u$. Say the amount of resources $P_v$ uses to help $P_u$ is $x$. When $P_u$ is done updating $v$'s subtree, the next $x$ time resources given to $P_u$ are passed to $P_v$. Notice that $P_v$ is guaranteed to receive those time tokens from $P_u$ being that every time a time resource is given to $P_v$ then a time resource is also given to $P_u$ as $u$ was an ancestor of $v$. This way, the subtree of $v$ is guaranteed to be completely updated with the label assigned to $v$ by $P_u$ before the new label assigned to $v_R$ by $P_v$ is even considered, and the processes are still guaranteed to complete on time.

\paragraph{When $w$ splits during $P_u$:}
In this case $P_u$ assigns a label for $u_R$. Then $w$ splits, and $P_w$ assigns a new label to $u_R$. At the end of the execution of both $P_w$ and $P_u$, the label assigned to $u_R$ by $P_w$ must be the label which is assigned in the appropriate locations in all of the tags of leaves in $u_R$'s subtree. This is guaranteed as follows. If $P_u$ has already updated the subtree of $u_R$ before $P_w$ reaches $u_R$'s subtree, then no special modifications need to be made. Also, in this case, it is not possible for $P_w$ to be in the process of updating $u_R$'s subtree when $P_u$ reaches $u_R$, as $P_u$ begins with $u_R$ on its first step. The only problematic situation in this case is if $P_u$ has already begun updating $u_R$'s subtree when $P_w$ reaches $u_R$'s subtree. In such a situation, $P_w$ uses its time resources to help $P_u$ finish updating $u_R$'s subtree. Say the amount of resources $P_w$ uses to help $P_u$ is $y$. Notice that,as opposed to the case considered above, $P_w$ cannot be guaranteed that any more insertions will be made into the subtrees of either $u_R$ or $u_L$, and therefore, the $y$ time resources that $P_w$ used to assist $P_u$ are cannot be guaranteed to be payed back by $P_u$, as $u$ isn't an ancestor of $w$. To solve this, $P_w$ performs double the work it would normally do (which is still $O(1)$) when assisting $P_u$ for each of the $y$ time resources, and then, for the next $y$ time resources which are given to $P_w$ after it is done assisting $P_u$, is also does double the work (this extra work will be going directly into updating the subtree of $u_R$ with the new label assigned to it by $P_w$).


\subsection{The Bottom Line}
To recap, each time a new element is added to $L$, there are $O(\log n)$ weight increases, where each weight increase might induce a split (which takes constant time), and also might perform a constant number of operations to update some labels and tags. Thus, each insertion requires $O(\log n)$ worst-case time. It is important to notice that due to the nature of the labeling and tagging scheme, it is possible to answer order queries correctly even while an update to the WBBT is taking place, as each tag-process does not create inconsistencies with other tag-processes. Moreover, the order of any two nodes can be decided by the order of the binary presentation of their tags, as opposed to using several tags per node, together with some timestamps. Thus, the following has been proven.

\begin{theorem}\label{thm:OLL}
It is possible to solve the MLLP with $O(\log n)$ worst-case relabels and time per insertion. Furthermore, order queries can still be answered correctly while tag-processing is taking place.
\end{theorem}

\section{Order-Maintenance Data Structure}\label{sec:OMP}
Indirection is used in order to achieve an $O(1)$ worst-case time bound per insertion, which closely follows the techniques of Dietz and Sleator in~\cite{DS87}. The list $L$ is partitioned into consecutive sublists. Each sublist is called a \textit{chunk}. The main idea follows from the following lemma.

\begin{lemma}[from~\cite{LO88,DS87}]\label{lem:split}
If every $k$ insertions into any chunk, the largest chunk is split into two roughly equally sized chunks, then the size of the largest chunk is $O(k\log n)$, and the total number of chunks is $O(\frac{n}{k})$
\end{lemma}

So if every $\log n$ insertions into any chunk a split takes place, the total number of sublists is $O(\frac{n}{\log n})$, and the size of each chunk is $O(\log^2n)$. In the following it is shown how to implement the chunks with the appropriate operations, and how the chunks are used in collaboration with the solution to the MLLP to efficiently solve the OMP.

\subsection{The Chunks}
The implementation of each chunk of size $O(\log^2n)$ is as follows. Each chunk is maintained with a tree of depth 2. The nodes in the sublist maintained by the chunk are the leaves of this tree which are all at depth 2. Each non-leaf other than possible the root has between $\frac{\log n}{2}$ and $\log n$ children. The root has at most $O(\log n)$ children. For every non-leaf node, the order of its $O(\log n)$ children is maintained using the averaging method as described above.

\full{
Due to space limitations, the exposition of the implementation is deferred to the full version. However, as mentioned, the techniques follow directly from Dietz and Sleator in~\cite{DS87}.

}
The following operations are needed on each chunk.

\subsubsection{Order Query} The order of any two leaves which are siblings can be determined from the tags given to those leaves by their parent, and the order of any two leaves with different parents can be determined by the tags of the parents given to them by the root. In any case this takes $O(1)$ time.

\subsubsection{Insertion} When a new node $u$ is inserted into a chunk, it will always be added after a node $v$ which was already in the chunk. Let $p$ be the parent of $v$ in the depth 2 tree, and let $r$ denote the root of this tree. At first, $u$ is inserted after $v$ in the order structure of $p$'s children (using the averaging method). If the insertion of $u$ increases the number of children of $p$ to be more than $\log n$, then $p$ is split into two by creating a new sibling called $p'$. This new sibling is inserted as a child of $r$, following $p$ in the order of the children of $r$. Notice that the number of children of $r$ is always bounded by $O(\log n)$. Starting from this point in time till $\frac{\log n}{4}$ insertions are made into the children of either $p$ or $p'$, each such insertion transfers the last two nodes from the children of $p$ to the children of $p'$. Then, during the next $\frac{\log n}{4}$ insertions into the children of either $p$ or $p'$, $p$ reassigns the labels of its children as follows. Using the same technique as in section~\ref{subsec:tag_process}, each label has an extra bit at the most significant location, and $p$ guarantees that the order is maintained during the reassignment using this bit (details are similar to those of Section~\ref{subsec:tag_process} and are thus omitted). In total, the entire process takes $O(1)$ time per insertion.

\subsubsection{Tracking the median} It will become apparent later that there is a need to track the median of each tree of depth 2 as insertions are made into that tree. This is done as follows. Let $m$ be the median of the chunk prior to the insertion, and in addition, maintain the number of nodes preceding $m$ and the number of nodes following $m$. When an insertion happens, the first step is to discover if this new node precedes or follows $m$. This is done with an order query. Then the appropriate counter is updated. If the counters differ by more than 1, then the median needs to move one step in the appropriate direction in the sublist in order to balance them out, updating the counters accordingly. This entire process takes $O(1)$ time per insertion.

\subsubsection{Locating the largest chunk} Locating the largest chunk is fairly standard and can be done using an auxiliary dynamic array of size $O(\log ^2n)$, where each entry in the array is a doubly linked list of all chunks of size equal to the index of that location. In addition, all of the non empty locations in this array are maintained in a doubly linked list. The key observation is that a size of a chunk can only change by 1 due to an insertion, and therefore, changes to this doubly linked list are very local. Details are standard and are thus omitted.

\subsubsection{Splitting around the median} Recall that every $\log n$ insertions into $L$, the largest chunk needs to split into two chunks of roughly equal size. This is done as follows. Let $m$ be the median of the chunk, let $p$ be its parent, and let $r$ be the root. Then $p$ needs to be split into two around $m$, and $r$ needs to be split into two around $p$, creating a new chunk. The splits are done using the same splitting method which is used during the insertion, so details are omitted. The total amount of time needed to perform this splitting is $O(\log n)$, and this process is spread over the next $\log n$ insertions made into $L$, for a total of $O(1)$ time per insertion. Notice that it is possible that a splitting process of $p$ due to it having too many children, as described during the insertion process, is happening concurrently with a splitting process of $p$ due to a chunk splitting around $m$. This situation can be solved by performing all operations twice as fast, completing the splitting process which began first, and only then proceeding to the next splitting process. This is similar to the techniques used in section~\ref{subsec:tag_process}, where colliding tag-processes pass their time resources to other tag-processes. Such a case still costs only $O(1)$ time per insertion.


\subsection{Combining Chunks with Monotonic List Labeling} As mentioned above, the list $L$ is partitioned into $O(\frac{n}{\log n})$ chunks, and a new chunk is created every $\log n$ insertions into $L$. In addition, a list of the roots of the chunks, ordered by $L$, is maintained via the solution presented in Section~\ref{sec:OLL} for the MLLP. Denote this ordered list of roots by $L_r$. The size of $L_r$ is $O(\frac{n}{\log n})$, and every $\log n$ insertions into $L$, one insertion is made into $L_r$ due to the largest chunk splitting. This process of inserting a new root into $L_r$ costs $O(\log n)$ time, and is spread over the following $\log n$ insertions made into $L$, before another insertion is made into $L_r$. Recall that the insertion into the monotonic list labeling structure can be done in parts without affecting order queries, due to Theorem~\ref{thm:OLL}. Thus the total time per insertion is $O(1)$ in the worst-case.

\subsection{Answering Order Queries} An $Order(u,v)$ query is answered as follows. First it needs to be established whether $u$ and $v$ are in the
same chunk or not. This is done in $O(1)$ time by checking if the root of the chunk of $u$ is the same as the root of the chunk of $v$.
If the chunks are the same, then the query is answered directly through the chunk. If not, then the query is answered by comparing the tags of
the roots of the chunks given by the monotonic list labeling structure. For simplicity sake, consider the tag of each node in $L$ to be the concatenation of the tag of its chunk representative in $L_r$, followed by the tags of its parent and itself within its chunk. This will simplify the explanations in Section~\ref{sec:POLP}.

Thus, the following has been proven.
\begin{theorem}\label{thm:OMP}
It is possible to solve the order maintenance problem where each operation costs $O(1)$ time in the worst-case.
\end{theorem}

\section{Adding Predecessor Queries}\label{sec:POLP}
In this section it will be shown how the POLP can be implemented within the bounds claimed. The results are summarized by Theorem~\ref{thm:polp} at the end of this section.

\subsection{y-fast-tries}
The y-fast-trie~\cite{Willard83} is picked as the predecessor data structure of choice in order to achieve the desired bounds, due to its simple presentation. It is possible to achieve the same bounds with other structures (such as the van-Emde Boas data structure~\cite{vEB77}). The y-fast-trie allows to answer predecessor queries over a set $S$ of size $m$ taken from universe $U$ in $O(\log \log u)$ time, where $u=|U|$. In addition, as will be shown, updates to $S$ (insertions and deletions) can be done in $O(\log \log u)$ expected time. The space usage is $O(m)$ words.

Being as the details of implementation of the y-fast-trie are of importance in the setting here, they are described briefly next. The y-fast-trie is based on another structure called the x-fast-trie, which is a trie of the binary presentations of elements in $S$ (so an edge to a left child corresponds to 0, while an edge to a right child corresponds to 1). Thus, the height of the x-fast-trie is $\log u$, and each node corresponds to a prefix of a binary presentation of some element (possibly more than one) in $S$. If a node only has a right child, then it maintains a pointer to the leaf with smallest key in the subtrie of its right child. Likewise, if a node only has a left child, then it maintains a pointer to the leaf with largest key in the subtrie of its left child. Finally, each node is maintained in a dynamic hash table, with the key being the binary prefix corresponding to the node (together with its length). Roughly speaking, a predecessor search on $x\in U$ performs a binary search on the binary presentation of $x$, and takes $O(\log \log u)$ worst-case time. An insertion of $x\in U$ is performed by inserting the new nodes corresponding to prefixes of the binary presentation of $x$ into the hash table, and possibly updating pointers from internal nodes to some leaves. This takes $O(\log u)$ worst-case expected time, where the expectation is due to the dynamic hash table. The space usage of the x-fast-trie is $O(m\log u)$ words.

Typically, the y-fast-trie uses the x-fast-trie as a top structure together with standard bucketing techniques. $S$ is partitioned into $O(\frac{m}{\log u})$ buckets, each bucket with $O(\log u)$ consecutive elements. Each bucket is maintained in a balanced binary search tree, BBST for short (AVL trees, or red-black trees). In addition, each bucket sends one representative to the x-fast-trie, which is now built on only $\frac{m}{\log u}$ elements, and so the space usage is now $O(m)$. An insertion is performed by inserting into the BBST of the appropriate bucket, and splitting the bucket if needed (causing an insertion to the x-fast-trie). A bucket is only split after $\Theta (\log u)$ insertions are made into that bucket, and so the time for inserting into the y-fast-trie is $O(\log \log u)$ amortized expected time. A query is performed by first querying the x-fast-trie, and then searching in the BBST of the appropriate bucket (with possible 1 or 2 more buckets near it) in $O(\log\log u)$ worst-case time.

However, the technique from Lemma~\ref{lem:split} can be used in order to make the insertion time worst-case expected by splitting the largest bucket every $\log n$ insertions. This way, $S$ is still partitioned into $O(\frac{m}{\log u})$ buckets, but each bucket has $O(\log^2 u)$ consecutive elements. Nevertheless, a predecessor search within a bucket still costs $O(\log \log u)$ time in the worst-case. An insertion into the x-fast-trie is now done as follows. Each node in the x-fast-trie is given a timestamp of when it was created. When the insertion process begins, the timestamp $\tau$ prior to the insertion is saved, and any predecessor query that is performed during the insertion process will ignore any data that has a timestamp after $\tau$. Once the insertion phase is completed, the structure is informed that it may ignore $\tau$ (or $\tau$ can be updated to the new timestamp after the insertion took place). Notice that the pointers to smallest or largest elements in some subtries also need to maintain these timestamps, and possibly another pointer to differentiate between the pointer prior to time $\tau$ and the pointer after time $\tau$. The $O(\log u)$ work needed to update the x-fast-trie is spread over the following $\Theta(\log u)$ insertions into any bucket, and finishes before another bucket splits. In addition, the splitting of the buckets is also done during those following $\Theta(\log u)$ insertions.

\subsection{Scheduling Splits of Buckets and Chunks}

Let $m=\sum_{i=1}^k |S_i|$. One option for solving the POLP is to maintain each of the sets $S_1,\cdots, S_k\subset L$ in a y-fast-trie, with the key of each element being its tag from the order-maintenance structure from Section~\ref{sec:OMP}. Notice that each key is contained within $O(\log n)$ bits, as the universe size of the tags is bounded by $n^c$ for some constant $c$. However, the problem with this solution is that each new element inserted into $L$ can cause a poly-logarithmic number of elements to change their tags (as it changes the tag of $O(1)$ chunks), which needs to be reflected by changing the keys in the y-fast-trie, and can be rather costly.

The first observation that can help solve this problem is to notice that the keys which need to be readjusted in the y-fast-trie are the keys of representatives of buckets, as the other elements are maintained in a BBST, and so their ordering never changes regardless of their tags. The second helpful observation is that while each insertion into $L$ causes many tags to change in the order-maintenance structure, only $O(1)$ tags of representatives of chunks in the monotonic list labeling structure are changed. Thus, the main idea is to guarantee that each chunk in the order-maintenance structure will contain 1 (possible 2 during a split) bucket representative from any of the y-fast-tries of any of the sets. However, care needs to be taken to guarantee that splitting process caused by buckets and chunks splitting do not interfere with each other. To this end, some more modifications are needed, as is described next.

\paragraph{Unifying splitting processes:} The first step in order to guarantee that each chunk has at most 1 bucket representative is to unify the splitting process over all the y-fast-tries of all of the sets. So now, every $\log n$ insertions into any of the y-fast-trie structures of any set, the largest bucket from all of the y-fast-tries is taken to be split. This guarantees, by Lemma~\ref{lem:split}, that only one y-fast-trie will be in a midst of an insertion process into its x-fast-trie component, while the size of any bucket is bounded by $O(\log^2 n)$. Furthermore, the total number of buckets in all the y-fast-trie structures is $O(\frac{n}{\log n})$, and so the total space used by all of the y-fast-tries is still linear.

\paragraph{Interfering splits:} Each time a new bucket representative is created, if the chunk which contains this representative already has a different representative within it, it will need to be split via a splitting process denoted by $P_b$. On the one hand this seems reasonable as such a process takes place only once every $\log n$ insertions into y-fast-trie structures, so its work can be spread over those insertions. However, it is possible that the order-maintenance structure is already in the midst of a chunk split process, denoted by $P_c$, due to its own machinery. Alternatively, it is possible that $P_c$ wants to begin while $P_b$ is currently executing.

This difficulty is solved as follows. The solution is shown for the first case (i.e.~$P_b$ begins while $P_c$ is in the midst of executing), as the second case simply reverses the roles of the processes. Going back to the terminology of section~\ref{subsec:tag_process}, each insertion into any of the y-fast-tries gives a time resource to $P_b$, while each insertion into $L$ gives a time resource to $P_c$. $P_b$ uses its time resources to help $P_c$ finish, but now $P_c$ needs to perform double the work per each time resource it receives. Let $x$ be the number of time resources passed from $P_b$ to $P_c$. When $P_c$ finally completes the current split, $P_b$ uses its next $x$ time resources to do double the work of what it normally would do, allowing it to catch up to where it would have been had the interference with $P_b$ not taken place. Notice that at most half the time resources given to $P_c$ will be passed on to $P_b$. Thus, each time resource is still used to perform $O(1)$ work.

\paragraph{Final run-time tuning:} Each insertion into either a y-fast-trie bucket or a chunk causes $O(1)$ representatives to change their tag. However, each such change of a tag will actually cost $O(\log n)$ worst-case expected time, as the binary presentation of the representative is changed, and this needs to be reflected in the x-fast-trie portion of the y-fast-trie. To solve this, the definitions of a chunk and bucket are slightly changed. Instead of creating a split every $\log n$ insertions, now a split is created every $\log^2 n$ insertions. Following Lemma~\ref{lem:split}, the size of the largest chunk or bucket is now bounded by $O(\log^3 n)$. The only additional change is that each chunk is implemented with a tree of depth 3 instead of a tree of depth 2. The rest of the details remain the same, up to some constants.

Now that a split occurs every $\log^2 n$ time resources, the scheduling is done as follows. Every $\log n$ time resources, $O(1)$ work is performed on the monotonic list labeling structure from Section~\ref{sec:OLL}. This causes $O(1)$ bucket representatives to change their tag, and so the $O(\log n)$ work needs to be performed in order to update the appropriate y-fast-tries with this tag change is spread over the next $\log n$ time resources. Thus each time resource pays for $O(1)$ work in expectation.

\paragraph{Running time:} The following has been proven.

\conf{
Each insertion into $L$ costs $O(1)$ expected time, while each insertion of an element from $L$ into a set $S$ will cost $O(\log \log n)$ expected time (due to inserting into the BBST of a bucket). A predecessor query on a set $S$ is performed as usual on a y-fast-trie, in $O(\log \log n)$ worst-case time. Order queries can still be answered in $O(1)$ worst-case time. Thus,
}

\begin{theorem}\label{thm:polp}
There exists an data structure for a dynamic ordered list $L$ of size $n$ and (disjoint) dynamic subsets $S_1,\cdots,S_k\subseteq L$ such that:(1) order queries are answered in $O(1)$ worst-case time, (2) inserting a node $u$ after a given node $v\in L$ takes $O(1)$ worst-case expected time, (3) for any $1\leq i \leq k$, inserting an element from $L/\bigcup_{j\neq i} S_j$ into $S_i$ takes $O(\log \log n)$ worst-case expected time, and (4) for any $1\leq i \leq k$ and $u\in L$, locating the predecessor of $u$ in $S_i$ takes $O(\log \log n)$ worst-case time.

\end{theorem}

\section{Acknowledgments}
The author would like to thank Neta Barkay, Michael Bender, Rajeev Raman, Moshe Lewenstein, and Milan Straka for useful discussions.

{\small
\bibliographystyle{alphainit}
\bibliography{tsvi}

\newcommand{\etalchar}[1]{$^{#1}$}
\begin{thebibliography}{BCD{\etalchar{+}}02}

\bibitem[AKLL05]{AKLL05}
A.~Amir, T.~Kopelowitz, M.~Lewenstein, and N.~Lewenstein.
\newblock Towards real-time suffix tree construction.
\newblock In {\em String Processing and Information Retrieval, 12th
  International Conference}, pages 67--78, 2005.

\bibitem[AN08]{AN08}
A.~Amir and I.~Nor.
\newblock Real-time indexing over fixed finite alphabets.
\newblock In {\em Proceedings of the Nineteenth Annual ACM-SIAM Symposium on
  Discrete Algorithms}, pages 1086--1095, 2008.

\bibitem[AV03]{AV03}
L.~Arge and J.~S. Vitter.
\newblock Optimal external memory interval management.
\newblock {\em SIAM J. Comput.}, 32(6):1488--1508, 2003.

\bibitem[BCD{\etalchar{+}}02]{BCDFZ02}
M.~A. Bender, R.~Cole, E.~D. Demaine, M.~Farach-Colton, and J.~Zito.
\newblock Two simplified algorithms for maintaining order in a list.
\newblock In {\em Algorithms - ESA 2002, 10th Annual European Symposium}, pages
  152--164, 2002.

\bibitem[BF02]{BF02}
P.~Beame and F.~E. Fich.
\newblock Optimal bounds for the predecessor problem and related problems.
\newblock {\em J. Comput. Syst. Sci.}, 65(1):38--72, 2002.

\bibitem[BI11]{BI11}
D.~Breslauer and G.~F. Italiano.
\newblock Near real-time suffix tree construction via the fringe marked
  ancestor problem.
\newblock In {\em String Processing and Information Retrieval, 18th
  International Symposium}, pages 156--167, 2011.

\bibitem[BKS12]{BkKS12}
J.~Bul{\'a}nek, M.~Kouck{\'y}, and M.~Saks.
\newblock Tight lower bounds for the online labeling problem.
\newblock In {\em Proceedings of the 44th Symposium on Theory of Computing
  Conference}, pages 1185--1198, 2012.

\bibitem[Die82]{Dietz82}
P.~F. Dietz.
\newblock Maintaining order in a linked list.
\newblock In {\em Proceedings of the 14th Annual ACM Symposium on Theory of
  Computing}, pages 122--127, 1982.

\bibitem[Die89]{Dietz89}
P.~F. Dietz.
\newblock Fully persistent arrays (extended array).
\newblock In {\em Algorithms and Data Structures, Workshop}, pages 67--74,
  1989.

\bibitem[DR93]{DR93}
P.~F. Dietz and R.~Raman.
\newblock Persistence, randomization and parallelization: On some combinatorial
  games and their applications (abstract).
\newblock In {\em Algorithms and Data Structures, Third Workshop, WADS '93,
  Montr{\'e}al, Canada,}, pages 289--301, 1993.

\bibitem[DS87]{DS87}
P.~F. Dietz and D.~D. Sleator.
\newblock Two algorithms for maintaining order in a list.
\newblock In {\em Proceedings of the 19th Annual ACM Symposium on Theory of
  Computing}, pages 365--372, 1987.

\bibitem[DSZ94]{DSZ94}
P.~F. Dietz, J.~I. Seiferas, and J.~Zhang.
\newblock A tight lower bound for on-line monotonic list labeling.
\newblock In {\em 4th Scandinavian Workshop on Algorithm Theory}, pages
  131--142, 1994.

\bibitem[FG04]{FG04}
G.~Franceschini and R.~Grossi.
\newblock A general technique for managing strings in comparison-driven data
  structures.
\newblock In {\em Automata, Languages and Programming: 31st International
  Colloquium}, pages 606--617, 2004.

\bibitem[KL07]{KL07}
T.~Kopelowitz and M.~Lewenstein.
\newblock Dynamic weighted ancestors.
\newblock In {\em 18th Annual ACM-SIAM Symposium on Discrete Algorithms,},
  pages 565--574, 2007.

\bibitem[LO88]{LO88}
C.~Levcopoulos and M.~H. Overmars.
\newblock A balanced search tree with {\it o } (1) worst-case update time.
\newblock {\em Acta Inf.}, 26(3):269--277, 1988.

\bibitem[PT06]{PT06}
M.~P\v{a}tra\c{s}cu and M.~Thorup.
\newblock Time-space trade-offs for predecessor search.
\newblock In {\em Proceedings of the 38th Annual ACM Symposium on Theory of
  Computing}, pages 232--240, 2006.

\bibitem[P\v11]{Patrascuprivate}
M.~P\v{a}tra\c{s}cu.
\newblock Private communication, 2011.

\bibitem[SKT{\etalchar{+}}00]{SGKTSB00}
A.~S. Szalay, P.~Z. Kunszt, A.~Thakar, J.~Gray, D.~R. Slutz, and R.~J. Brunner.
\newblock Designing and mining multi-terabyte astronomy archives: The sloan
  digital sky survey.
\newblock In {\em Proceedings of the 2000 ACM SIGMOD International Conference
  on Management of Data}, pages 451--462, 2000.

\bibitem[Ukk95]{Ukkonen95}
E.~Ukkonen.
\newblock On-line construction of suffix trees.
\newblock {\em Algorithmica}, 14(3):249--260, 1995.

\bibitem[vEB77]{vEB77}
P.~van Emde~Boas.
\newblock Preserving order in a forest in less than logarithmic time and linear
  space.
\newblock {\em Inf. Process. Lett.}, 6(3):80--82, 1977.

\bibitem[Wei73]{Weiner73}
P.~Weiner.
\newblock Linear pattern matching algorithms.
\newblock In {\em 14th Annual Symposium on Switching and Automata Theory},
  pages 1--11, 1973.

\bibitem[Wil83]{Willard83}
D.~E. Willard.
\newblock Log-logarithmic worst-case range queries are possible in space
  theta(n).
\newblock {\em Inf. Process. Lett.}, 17(2):81--84, 1983.

\bibitem[Wil86]{Willard86}
D.~E. Willard.
\newblock Good worst-case algorithms for inserting and deleting records in
  dense sequential files.
\newblock In {\em Proceedings of the 1986 ACM SIGMOD International Conference
  on Management of Data}, pages 251--260, 1986.

\end{thebibliography}
}

\end{document}